\documentclass[usenatbib]{mn2e}
\usepackage{epsfig}
\usepackage{amsmath}
\usepackage{ulem}

\def\gsim{\mathrel{\raise0.35ex\hbox{$\scriptstyle >$}\kern-0.6em 
\lower0.40ex\hbox{{$\scriptstyle \sim$}}}}
\def\lsim{\mathrel{\raise0.35ex\hbox{$\scriptstyle <$}\kern-0.6em 
\lower0.40ex\hbox{{$\scriptstyle \sim$}}}}
\def\gs{\mathrel{\raise0.35ex\hbox{$\scriptstyle >$}\kern-0.6em 
\lower0.40ex\hbox{{$\scriptstyle \sim$}}}}
\def\ls{\mathrel{\raise0.35ex\hbox{$\scriptstyle <$}\kern-0.6em 
\lower0.40ex\hbox{{$\scriptstyle \sim$}}}}

\def\lesssim{\mathrel{\hbox{\rlap{\hbox{\lower4pt\hbox{$\sim$}}}\hbox{$<$}}}}
\def\gtrsim{\mathrel{\hbox{\rlap{\hbox{\lower4pt\hbox{$\sim$}}}\hbox{$>$}}}}

\date{\today}
\title[Proto-Clusters at $z\sim2.5$]
{Proto-Clusters with Evolved Populations around Radio Galaxies at $z\sim2.5$}

\author[M. Kajisawa et al.]{
\parbox[t]{\textwidth}{
Masaru Kajisawa$^{1}$\thanks{E-mail: kajisawa@optik.mtk.nao.ac.jp},
Tadayuki Kodama$^{1}$,
Ichi Tanaka$^{2}$,
Toru Yamada$^{2}$,\\
Richard Bower$^{3}$
}
\vspace*{6pt}\\
$^{1}$National Astronomical Observatory of Japan, Mitaka, Tokyo 181--8588, Japan\\
$^{2}$Subaru Telescope, National Astronomical Observatory of Japan, 650 North Aohoku Place, Hilo, HI 96720, USA\\
$^{3}$Department of Physics, University of Durham, South Road, Durham DH1 3LE, UK
}

\begin{document}

\maketitle

\begin{abstract}
We report a discovery of proto-cluster candidates around high redshift
radio galaxies at $z\sim2.5$ on the basis of clear statistical
excess of colour-selected galaxies around them seen in the deep near-infrared
imaging data obtained with CISCO on Subaru Telescope.  
We have observed six targets, all at similar redshifts at $z\sim2.5$,
and our data reach to $J=23.5$, $H=22.6$ and $K=21.8$ (5$\sigma$) and
cover a $1.6' \times 1.6'$ field centered on each radio galaxy.
We apply colour cuts in $JHKs$ in order to exclusively search for 
galaxies located at high redshifts, $z>2$.  Over the magnitude range
of $19.5<K<21.5$ we see a significant excess of red galaxies with
$J-K>2.3$ by a factor of two around the combined radio galaxies 
fields compared to those found in the general field of GOODS South.
The excess of galaxies around the radio galaxies fields 
becomes more than factor of three around $19.5<K<20.5$ 
when the two-colours cuts are applied with $JHKs$.
Such overdensity of the colour-selected galaxies suggest that those
fields tend to host 
high density regions at high redshifts, although there seems to be the
variety of the density of the colour-selected galaxies in each field.
In particular, two radio galaxies fields out of the six observed
fields show very strong density 
excess and these are likely to be proto-clusters
associated to the radio 
galaxies which would evolve into rich clusters of galaxies dominated
by old passively evolving galaxies.
\end{abstract}


%
%
\section{Introduction}

\label{sec:intro}

From many previous studies based on fundamental planes and
colour-magnitude (CM) relations in clusters up to $z\sim 1.3$,
it appears that major star formation in the brightest cluster
galaxies was virtually completed by $z\sim 2$ (e.g., 
 \citealp{bow92}, \citealp{ell97}, \citealp{kod98}, \citealp{sta98},
 \citealp{van98}, \citealp{bla03}).
Therefore, one expects to find proto-clusters dominated by
already passively evolving galaxies even beyond $z\sim 2$.
Searching for such ancestors of present-day rich clusters of galaxies
is not an easy task, however, since it requires a large volume to
survey in order to find such rare objects.
Radio galaxies can be viewed as good markers of such high density
regions at high redshifts, since they are among the most massive galaxies
at any redshift ($M_{\rm star}>10^{11} M_{\odot}$, \citealp{roc04}).
In fact, about half of powerful radio galaxies at intermediate
redshift inhabit rich environments (\citealp{hil91}, \citealp{bor06}).
The hypothesis that radio galaxies point us towards proto-clusters
at high redshift has been further supported by the recent finding of a
correlation between AGN activity and bulge mass (Magorrian relation,
\citealp{geb00}). 
Given this relation, it is natural to imagine that massive cD galaxies
sitting in deep potential well of clusters tend to host massive powerful AGN
and hence they are viewed as radio galaxies at high redshift.

Narrow-band surveys of Ly$\alpha$ emitters around high-$z$ radio galaxies
have been intensively conducted by the Leiden group (e.g.,
\citealp{kur00}, \citealp{ven02}) 
over many years who discovered a large number of emitters at the same
redshifts of the radio galaxies, and a large fraction of those emitters have
been spectroscopically confirmed (e.g., \citealp{kur00}, \citealp{ven02}).
These overdense regions of star forming galaxies are embedded in large scale
structures at high redshifts and are likely to evolve into massive systems
at the present day, such as clusters of galaxies.

Alternative approach is to find evolved populations at longer wavelength.
This is complementary to the emitter surveys.
\cite{bes03} found an excess of red galaxies with $R-K>4$ around
radio-loud AGNs at $z\sim1.6$.
\cite{hal98} and \cite{kod03} also found statistical overdensities of red
galaxies around radio-loud AGNs at $1\lsim z\lsim1.5$ in near-infrared surveys.
\cite{wol03} also report overdensities of red galaxies in optical--NIR colours
around 13 QSO's at $2\lsim z\lsim3$.
These studies indicate that there is also an excess of {\it evolved} population
around the high-$z$ radio galaxies, as expected if they are embedded in proto-clusters.

We take the second approach and extend the previous analyses to higher
redshift based on the near-infrared (NIR) imaging. 
And in this {\it Letter}, we report a discovery of the overdensity of NIR-selected
galaxies in the fields of six radio galaxies at $z\sim2.5$.

The Vega-referred magnitude system is used throughout the paper.
%
%
\section{Observation and Data Reduction}
\label{sec:obs}
We performed the $JHK$-bands imaging observations of the fields of 
the five radio galaxies and one radio-loud QSO (4C +04.81) at $z\sim2.5$ 
with CISCO \citep{mot02} on the Subaru Telescope  
on 2005 March 21-22 and September 19-20. CISCO has the field of view of
about 1.8 $\times$ 1.8 arcmin$^{2}$ with 0.105'' pixel scale. 
The target information and
exposure time are given in Table \ref{tab:obs}.
The weather conditions were stable during the observations, and seeing
sizes were between 0.3'' and 0.7'' (FWHM). 

The data were reduced using the IRAF software package as described in
\cite{kaj05}.
We performed flat-fielding with the superflat frames for the
CISCO data \citep{mot02}, and subtracted the dark and sky background.
Then the data were co-registered and combined.
We also reduced the archival CISCO data for 4C -00.62 (5600 sec in
$J$-band and 1040 sec in $K$-band) and 4C 
+23.56 (5600 sec in $J$-band, 1600 sec in $H$-band and 1200 sec in 
$K$-band) and combined together.
These are included in Table \ref{tab:obs}.
Our final multi-band combined images have the slightly smaller field
of view of about 1.6 $\times$ 1.6 arcmin$^{2}$, which corresponds to 
$\sim0.77 h_{70}$Mpc on a side at $z\sim2.5$ (We assume a cosmology of 
$\Omega_{0}=0.3$, $\Omega_{\Lambda}=0.7$). 
For each target, the images were convolved with a Gaussian kernel 
to match the PSFs to the worst one.
The final PSF for each target is also given in Table \ref{tab:obs}.
\begin{table}
 \centering
 \begin{minipage}{140mm}
  \caption{Summary of the observational data}
  \label{tab:obs}
  \begin{tabular}{@{}lcccccc@{}}
  \hline
   Name     & Redshift & J & H & Kp & PSF &\\
            &          &(sec)&(sec)&(sec)&(arcsec)&\\ 
 \hline
 4C -00.62 & 2.527 & 6560 & 2220 & 3880 & 0.67 &\\ 
 4C -00.54 & 2.363 & 5760 & 3200 & 3200 & 0.65 & \\
 4C +23.56 & 2.483 & 6800 & 4525 & 2640 & 0.67 & \\
 MRC 2104-242 & 2.419 & 5400 & 3630 & 2880 & 0.60 & \\
 4C +04.81 & 2.594 & 5400 & 4320 & 2880 & 0.45 & \\
 MRC 0406-244 & 2.440 & 5400 & 3960 & 2560 & 0.45 & \\
\hline
\end{tabular}
\end{minipage}
\end{table}

Source detection was performed in the K-band images using the
SExtractor image analysis package \citep{ber96}. 
We adopted MAG\_AUTO from the SExtractor as the total $K$-band
magnitude of detected objects. For the colour measurements, 
we used the MAG\_AUTO with the Kron factor of $k=2$, which corresponds
to the 20\% smaller aperture size than that used in total magnitude
with default value $k=2.5$. Kron radius of each object was determined
in the $K$-band images and the same radii were used for all $J$, $H$,
and $K$-bands.

The UKIRT Faint Standards were used for the flux calibration.
For the data obtained in the March run, we had to shift the $H$-band
zero-point by 0.13 mag in order to reproduce the stellar tracks in $JHK$
\citep{kid03} using the observed stellar objects in our combined images.

The 5$\sigma$ limiting magnitude at 1.2 arcsec diameter is about 
 $J=23.5$, $H=22.6$ and $K=21.8$, respectively.
All the magnitudes are corrected for the Galactic extinction, which is
estimated at the positions of the radio galaxies (near the center of each
observed field) based on \cite{sch98}.
%
%
\section{Colour selection of high-$z$ galaxies}
\label{sec:colsel}
\begin{figure*}
\begin{center}
\leavevmode
\epsfxsize 0.9\hsize \epsfbox{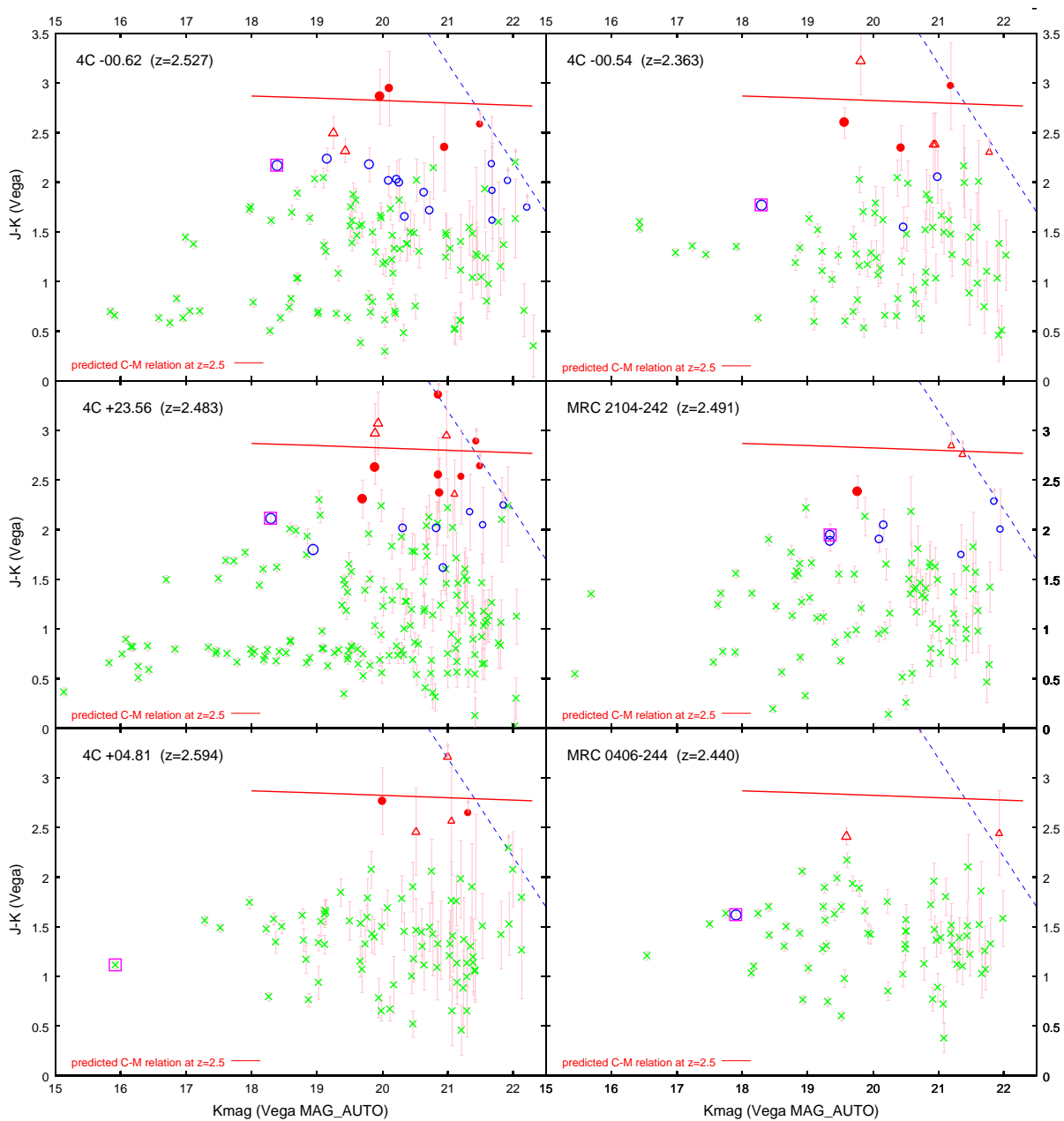}
\end{center}
\caption{Colour-magnitude diagrams of the six observed radio galaxies
fields. 
Type of the symbols differenciate the adopted colour selections as indicated
in Figure 2 (see also text).
Dashed lines show the 2$\sigma$ limit at $J$-band.
For objects with $J$-band fluxes lower than the limit, 
the lower limits of $J-K$ color are plotted by symbols with arrows. 
A square in each panel indicates the targeted radio galaxy.
Solid lines represent the expected C-M relation at $z=2.5$ calculated
from a GALAXEV single burst model with a duration of 0.1Gyr 
with $z_{\rm form}=5$.
The metallicity of the model is adjusted so that the model coincide with 
the C-M relation of Coma cluster \citep{bow92} at $z=0.02$.}
\label{fig:jk}
\end{figure*}

Figure \ref{fig:jk} shows $J-K$ vs $K$ colour-magnitude diagrams of
the six
observed radio galaxies fields. Type of symbols differenciate the galaxies
selected with different colour cuts on the $J-K$ vs. $H-K$ diagram
(Figure \ref{fig:jhk}, see below).
The square in each diagram indicates the targeted radio galaxy
(radio-loud QSO) at $z\sim2.5$. 
Some of the fields contain a bunch of very red galaxies ($J-K\gsim2.3$),
in particular 4C~$-$00.62 and 4C~+23.56 show sequences of red galaxies
which are likely to be progenitors of the colour-magnitude relation
of early-type galaxies commonly seen in lower redshift clusters at
$z\lsim1$. 
Therefore these two fields are most plausible high-$z$ cluster
candidates.
For reference, we also plotted an expected C-M relation at $z=2.5$ 
using the GALAXEV population synthesis spectral library \citep{bru03}.
We assumed a single burst with a duration of 0.1Gyr 
and the formation redshift of $z_{\rm
form}=5$. 
The observed red galaxies tend to be relatively bluer than 
the model prediction.
In contrast, the MRC 0406-244 field contain few red galaxies, which may
indicate the variety of the number densities of red companions around
radio-loud AGNs at $z\sim2.5$.

\begin{figure*}
\begin{center}
\leavevmode
\epsfxsize 0.9\hsize \epsfbox{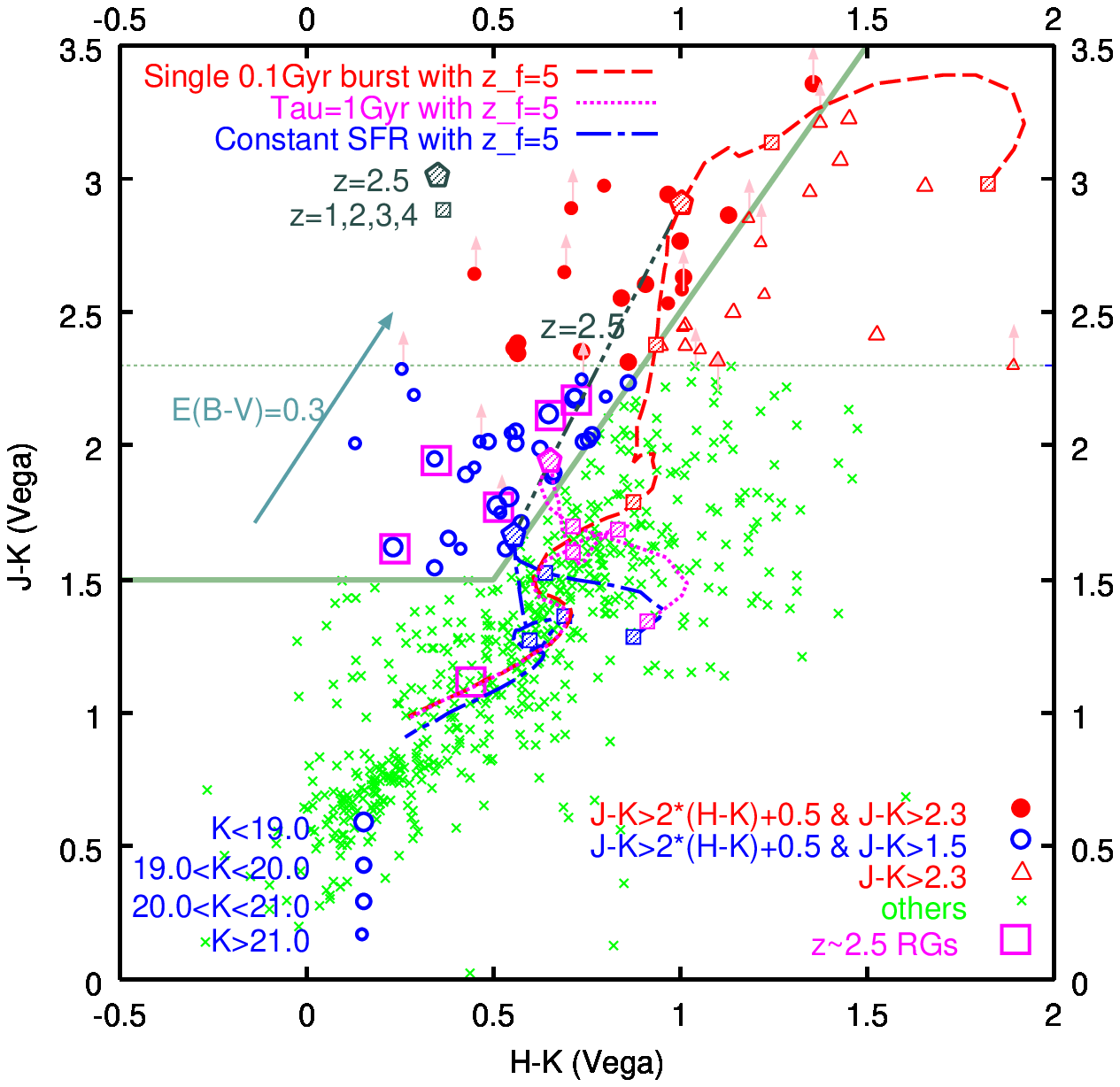}
\end{center}
\caption{Combined $J-K$ vs. $H-K$ colour-colour diagram of all the six radio
galaxies fields.
The horizontal dotted line and the thick solid lines show the boundaries of
our single colour selection of $J-K>2.3$ and the two-colours selection
with $JHK$, respectively.
Different symbols (solid circle, open circle, triangle, cross) 
correspond to our different colour selections as shown in the legend
(see also text). 
The size of the symbols for the colour-selected galaxies 
is scaled according to apparent magnitudes in $K$-band
(the bigger, the brighter). 
The six open squares indicate the targeted radio galaxies at $z\sim2.5$.
The long arrow at the left side of the figure shows a reddening vector of
Calzetti et al.(2000)'s law of 
E$(B-V)=0.3$ at $z=2.5$.
Dashed, dotted and dotted-dash curves represent the GALAXEV models of
evolutionary 
sequences over $0<z<4$ 
of galaxies with $z_{\rm form}=5$ for various star formation
histories.
Pentagons show the points of $z=2.5$ in the model sequences and shaded 
squares show those of $z=$1,2,3, and 4.
}
\label{fig:jhk}
\end{figure*}
In order to efficiently pick up plausible proto-cluster members associated
to the central radio galaxies at $z\sim2.5$, we make use of our multi-colour
photometric data at NIR.
We apply three different colour cuts as shown in Figure \ref{fig:jhk},
a combined colour-colour diagram in $J-K$ vs. $H-K$ for all the six fields.

First is a simple and single colour cut of
\begin{equation}
J-K>2.3
\end{equation}
as shown in the horizontal dotted line.
This colour-cut is the same as is used for the selection of the
Distant Red Galaxies 
at $z\gsim2$, and many of thus selected galaxies are spectroscopically
confirmed. 
(e.g., \citealp{van03}, \citealp{for04}, \citealp{red05}).
In fact, such red colours of galaxies can only be reproduced either by
Balmer/4000\AA-break galaxies with old populations at $z\gsim2$ or the
foreground galaxies with heavy dust extinction.

Second colour cut uses two colours as:
\begin{equation}
J-K > 2\times(H-K)+0.5\ \&\ J-K > 1.5 
\end{equation}
which are shown by thick solid lines in the figure.
Here we overplot the model colour tracks with various star formation histories
using the GALAXEV library.
As shown, the top left corner of the diagram separated by Equation (2)
should be exclusively populated by galaxies located at $2\lsim z\lsim3$.
This criterion is met while the Balmer/4000\AA\-break of galaxies falls between
$J$-band and $H$-band, while most of the foreground and background galaxies
would not meet such criterion.  Therefore, by applying such colour-cut
in $JHK$, 
we can effectively achieve strong contrast of plausible proto-cluster members
at $z\sim2.5$ against numerous foreground/background contamination. 
This two-colour-based selection also has a significant advantage over the first
one based on the single $J-K$ colour, since the two-colours cut can include
younger galaxies or star forming galaxies at $z\sim2.5$ which have
relatively blue 
colours in $J-K$ and would have been missed by the single $J-K>2.3$ cut.
It should be also noted that this selection is robust against dust extinction,
since the reddening vector (the long arrow in Figure \ref{fig:jhk}) is
almost parallel 
to the boundary line of the $JHK$-selection.
We also confirmed with the \cite{kna04}'s star catalogue that most of
cool M, L and 
T dwarfs should be excluded by this criterion, since they are either
too blue in $J-K$ 
or located on the right side of the boundary line of the $JHK$-selection.

\begin{table}
 \centering
  \caption{Number densities of the colour-selected galaxies with
  $K<21.8$ (arcmin$^{-2}$)}
  \label{tab:num}
  \begin{tabular}{@{}lcccc@{}}
  \hline
   Field    & $J$$-$$K$$>$2.3 & $JHK$-selected & $JHK$-selected & \\
            &                 &              & \& $J$$-$$K$$>$2.3 & \\ 
 \hline
 4C~$-$00.62    & 2.28 $\pm$ 0.93 & 6.08 $\pm$ 1.52 & 1.52 $\pm$ 0.76 & \\ 
 4C~$-$00.54    & 2.78 $\pm$ 1.05 & 2.38 $\pm$ 0.97 & 1.19 $\pm$ 0.69 & \\
 4C~+23.56      & 5.53 $\pm$ 1.48 & 6.72 $\pm$ 1.63 & 3.95 $\pm$ 1.25 & \\
 MRC~2104$-$242 & 1.15 $\pm$ 0.66 & 2.30 $\pm$ 0.94 & 0.38 $\pm$ 0.38 & \\
 4C~+04.81      & 2.31 $\pm$ 0.95 & 1.16 $\pm$ 0.67 & 1.16 $\pm$ 0.67 & \\
 MRC~0406$-$244 & 0.38 $\pm$ 0.38 & 0.38 $\pm$ 0.38 & 0.00 $\pm$ 0.00 & \\
\hline
-0062 \& +23.56 & 3.87 $\pm$ 0.87 & 6.97 $\pm$ 1.16 & 2.71 $\pm$ 0.72
  & \\
the other 4 fields & 1.68 $\pm$ 0.39 & 1.68 $\pm$ 0.39 & 0.65 $\pm$
  0.25 & \\
\hline
all 6 fields & 2.39 $\pm$ 0.39 & 3.39 $\pm$ 0.46 & 1.32 $\pm$ 0.29 & \\
\hline
\hline
 GOODS$-$S      & 1.66 $\pm$ 0.97 & 1.75 $\pm$ 1.03 & 0.66 $\pm$ 0.60 & \\
\hline
\end{tabular}
\end{table}

Different types of the symbols correspond to different colour selections.
Filled circles show the galaxies which satisfy the both criteria
simultaneously. 
Open circles represent the objects which meet only the second criterion,
and open triangle indicate those which meet only the first one
(i.e., DRGs which do not satisfy the second colour-cut hence are likely
to be background red objects).  All the others are shown by crosses.

It is notable that all the five radio galaxies at $z\sim2.5$ (squares in Figure
\ref{fig:jhk}) satisfy the two-colours cut (the remaining square outside of the
top left corner is the radio-loud QSO 4C~+04.81).
This supports that our colour selection works
reasonably well, even though the strong emission lines from the AGN component
could affect their colours \citep{iwa03}.

\begin{figure*}
\begin{center}
\leavevmode
\epsfxsize 0.95\hsize \epsfbox{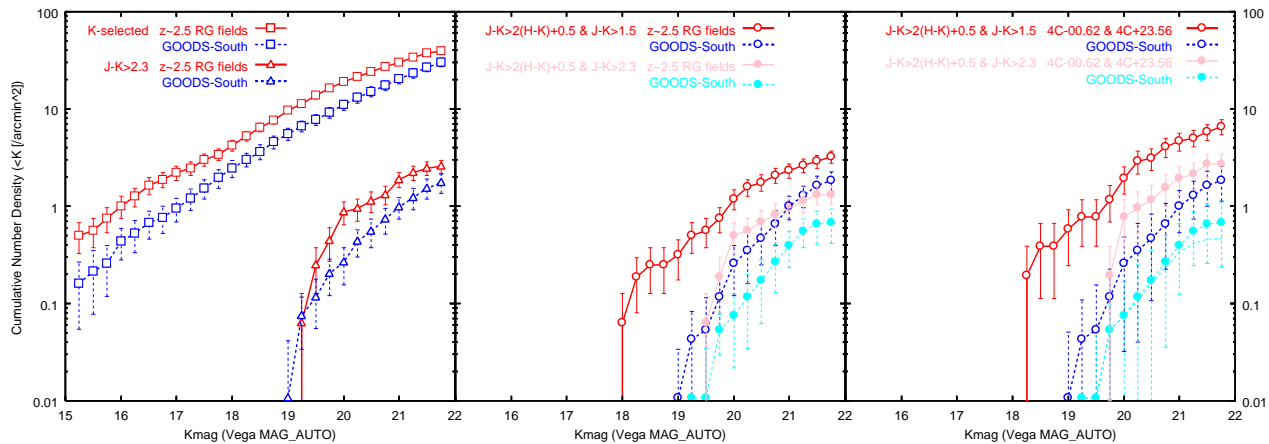}
\end{center}
\caption{Combined cumulative number counts of galaxies in the radio galaxies
fields (solid lines) and those in a controle field GOODS-South (dashed lines).
{\bf Left:} For all the $K$-selected objects (squares) and for red
galaxies with $J-K>2.3$ (triangles).
{\bf Middle:} For galaxies selected by two-colours cut (open circles)
and for galaxies which satisfy the two criteria simultaneously (solid circles).
{\bf Right:} Same as the middle panel but only for the two most probable proto-clusters
of 4C~$-$00.62 and 4C~+23.56.
Errorbars on the radio galaxies fields are Poissonian (i.e., square
roots of the observed numbers), and 
errorbars on the GOODS-S data indicate the field-to-field variance 
scaled to the same area as the radio galaxies fields 
(six, six and two times 1.6 $\times$ 1.6 arcmin$^{2}$ for 
the left, middle and the right panel, respectively).
}
\label{fig:nc}
\end{figure*}
Table \ref{tab:num} list the number densities of the colour-selected galaxies
with $K<21.8$ in each radio galaxy field. 
Poisson errors of these densities are also shown.
In addition, we used the public GOODS-South data 
(see next section for the detail) to estimate the degree of the 
field-to-field 
variance for the observed area of each radio galaxy field.
A random field of 1.6$\times$1.6 arcmin$^{2}$ was selected from
GOODS-S field and 
the numbers of color-selected galaxies were counted. We repeated this
 procedure and measured the one sigma variance in the numbers.
Such field-to-field variance is shown as 
errors in the number densities for the GOODS-S field in Table
\ref{tab:num} and Figure \ref{fig:nc} (see next section).

It is clear that the two fields, 4C~$-$00.62 and 4C~+23.56, contain quite
high number densities of galaxies that satisfy the two-colours cut (Eq.\ 2)
and they are among the most plausible proto-cluster candidates.
In particular, the 4C~+23.56 field has numerous red galaxies with $J-K>2.3$.
In fact, \cite{kno97} also reported a high density of red galaxies with $I-K>4$
in the central 1.25 arcmin$^{2}$ field around 4C~+23.56.
On the other hand, MRC 0406-244 field has few colour-selected galaxy.
There seems to be the variety of the density of the
colour-selected galaxies in each radio galaxy field.
In the following section, we compare those number densities of these
colour-selected 
galaxies in the radio galaxies fields with those in a general blank field.

%
%
\section{Overdensities of high-$z$ galaxies}
\label{sec:overdense}
In order to quantify the overdensities of the colour-selected galaxies around
the high-$z$ radio galaxies, we use GOODS-South survey \citep{gia04} as a
comparison field for which the public VLT/ISAAC ver1.5 data are available.

The $JHK$-bands data of the GOODS-S has similar depths and similar PSF
to those of our CISCO data, and are suitable for our purpose.
The area used in the analysis is 94 arcmin$^{2}$, which is
significantly larger than that of our radio galaxies fields.
We performed the source detection and photometry exactly in the same
manner as for the CISCO images.

In Figure \ref{fig:nc}, we compare the cumulative $K$-band number densities of 
the colour-selected galaxies in the radio galaxies fields and in the GOODS-S field.
In the left panel, we plot the number counts of all the $K$-selected objects
and the galaxies with $J-K>2.3$.
Note that the excess of all the $K$-selected objects at the 
bright end ($K\lesssim19$) is due to stars in the radio galaxies fields,
especially in the case of 4C~+23.56 field which contain numerous stellar objects
(Figure \ref{fig:jk}). 
The number counts of galaxies that are not selected by any color cut (``the
others'' in Figure \ref{fig:jhk}) in the radio galaxies fields and GOODS-S
are consistent, if we exclude these star excess in the radio galaxies
fields by the selection such as $J-K>1.0$ or $K>19$.
The excess of all the $K$-selected objects at $K>19$ is caused by the 
overdensity of the color-selected galaxies as seen below.

\begin{figure*}
\begin{center}
\leavevmode
\epsfxsize 0.75\hsize \epsfbox{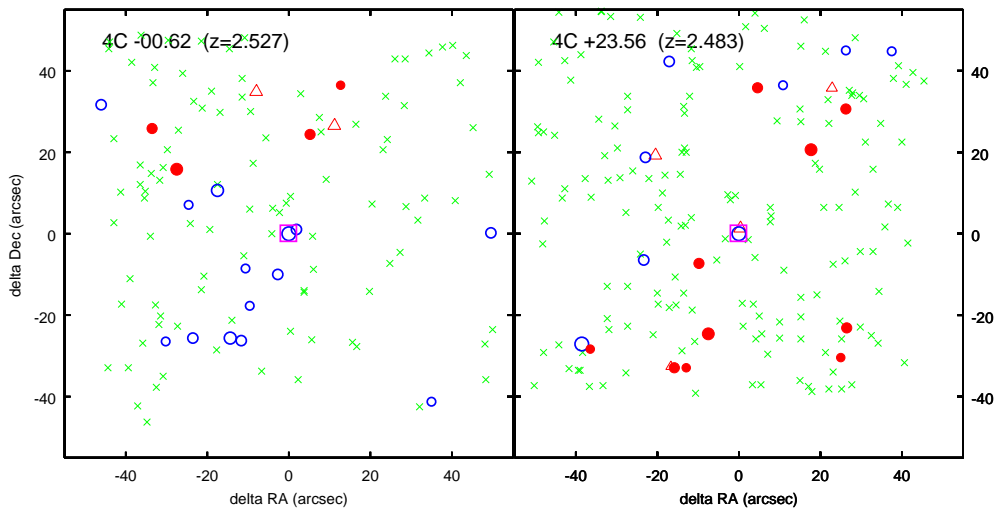}
\end{center}
\caption{Spatial distribution of the two most probable proto-clusters, 
4C -00.62 and 4C +23.56. Symbols are the same as Figure \ref{fig:jhk}.
}
\label{fig:xy}
\end{figure*}
The cumulative number density of the galaxies with $J-K>2.3$ in all the six radio
galaxies fields significantly exceeds that in the GOODS-S at $K\gtrsim19.5$.
The former density is twice higher than the latter over the magnitude range of
$19.5<K<21.5$.

The middle panel of Figure \ref{fig:nc} shows the cases for galaxies
selected by the two-colours cut in $JHK$ (open circles).
The galaxies shown by the filled circles have been further selected
as having red colours of $J-K>2.3$ in addition to the two-colours cut.
We see a significant excess in densities of the $JHK$-selected
galaxies around the radio galaxies.
In fact, the overdensity is more than factor of three compared to that
of the GOODS-S at $19.5<K<20.5$.
Red subsample with $J-K>2.3$ also shows similar overdensity of about
factor three over the range of $19.5<K<22$.


In the right panel of Figure \ref{fig:nc}, we show the same plots as
in the middle panel but for only the two highly plausible proto-cluster
candidates, namely 4C~$-$00.62 and 4C~+23.56 fields. 
The overdensity of galaxies selected by the two-colours cut is now
more than factor of five compared to the general field. 
The excess is even higher for the red subsample with $J-K>2.3$.
Such strong excess of NIR-selected (approximately stellar mass-selected)
galaxies indicates that these fields are likely to be 
proto-clusters with evolved populations associated to the central radio
galaxies at $z\sim2.5$, which would evolve into rich clusters of galaxies
today dominated by old passively evolving galaxies.
In fact, these two fields mainly cause the significant excess seen in
all the six fields, while the average density of the other four 
fields is consistent with that of the general field within the uncertainty
 (see Table \ref{tab:num}).

In Figure \ref{fig:xy}, we show the spatial distribution of
the colour-selected galaxies in the two proto-cluster candidates around the
$z\sim2.5$ radio galaxies.
Despite of poor statistics, it is suggestive that those colour-selected
member candidates are distributed non-uniformly around the central
radio galaxies.  
Such non-uniform distribution of galaxies in proto-clusters has also
been reported 
recently by \cite{cro05} who found a filamentary structure in the spatial
distribution of spectroscopically confirmed Ly$\alpha$ emitters associated to
a radio galaxy at $z=2.16$.
The mass-selected members of proto-clusters may similarly trace such
filamentary 
structures at least over the scale of $\sim0.8 h_{70}$Mpc at this
observed epoch. 
\section{Summary}
\label{sec:summary}
We searched for proto-clusters around the radio galaxies at $z\sim2.5$
based on the deep $JHK$-bands imaging taken with Subaru telescope.
In order to largely subtract foreground/background contamination and
have high contrast of proto-cluster member candidates associated to
the radio galaxies, we applied several colour selections;
(1) a widely used single colour cut of $J-K>2.3$ and (2) our new two-colours
cut with $JHK$-bands.  In both cases, we see a clear excess in the number
densities of colour-selected galaxies in the observed radio galaxies fields
compared to the general field GOODS-South. 
The overdensity in the six combined radio galaxies fields is about factor
of two for the galaxies with $J-K>2.3$ and is factor of three for those
selected by the two-colours cut, although the variety of the
overdensity in each field is also seen.  
In particular, two radio galaxies fields,
namely 4C~$-$00.62 and 4C~+23.56, show the highest overdensities of the
colour-selected galaxies of more than factor of five. 
Therefore, we are likely to be witnessing the assembly and formation of
proto-clusters with many evolved galaxies at $z\sim2.5$ associated to the
radio galaxies.
%
%
\section*{Acknowledgements}
We thank Dr.\ Kentaro Aoki who supported our CISCO observations.
We thank Drs.\ Carlos de Breuk, Jo\"el Vernet, Nick Seymour, 
Margrethe Wold and Alvio Renzini for discussion on the target selection
and the results.
This work was financially supported in part by a Grant-in-Aid for the
Scientific Research (No.\, 15740126) by the Japanese Ministry of Education,
Culture, Sports and Science.
This study is based on data collected at Subaru Telescope, which is operated by
the National Astronomical Observatory of Japan.
This study is also in part based on data collected at Very Large
Telescope at the ESO Paranal Observatory under Program ID: LP168.A-0485. 
Data reduction/analysis was carried out on ``sb'' computer system
operated by the Astronomical Data Analysis Center (ADAC) and Subaru
Telescope of the National Astronomical Observatory of Japan.
%
%

\end{document}